\documentclass[12pt]{article}

\usepackage{epsf,amsfonts,hyperref}
\usepackage{cite}
\newcommand{\newsection}{    
\setcounter{equation}{0}
\section}
\renewcommand{\appendix}[1]{
    \addtocounter{section}{1}
    \setcounter{equation}{0}
    \renewcommand{\thesection}{\Alph{section}}
    \newsection*{Appendix \thesection\protect\indent #1}
    \addcontentsline{toc}{section}{Appendix \thesection\ \ \ #1}
}
\newcommand\encadremath[1]{\vbox{\hrule\hbox{\vrule\kern8pt
\vbox{\kern8pt \hbox{$\displaystyle #1$}\kern8pt}
\kern8pt\vrule}\hrule}}
\def\enca#1{\vbox{\hrule\hbox{
\vrule\kern8pt\vbox{\kern8pt \hbox{$\displaystyle #1$}
\kern8pt} \kern8pt\vrule}\hrule}}

\newcommand\figureframex[3]{
\begin{figure}[bth]
\hrule\hbox{\vrule\kern8pt
\vbox{\kern8pt \vbox{
\begin{center}
{\mbox{\epsfxsize=#1.truecm\epsfbox{#2}}}
\end{center}
\caption{#3}
}\kern8pt}
\kern8pt\vrule}\hrule
\end{figure}
}
\newcommand\figureframey[3]{
\begin{figure}[bth]
\hrule\hbox{\vrule\kern8pt
\vbox{\kern8pt \vbox{
\begin{center}
{\mbox{\epsfysize=#1.truecm\epsfbox{#2}}}
\end{center}
\caption{#3}
}\kern8pt}
\kern8pt\vrule}\hrule
\end{figure}
}

\newcommand{\eq}[1]{eq.~(\ref{#1})}

\newcommand{\beq}{\begin{equation}}
\newcommand{\eeq}{\end{equation}}
\newcommand{\bea}{\begin{eqnarray}}
\newcommand{\eea}{\end{eqnarray}}

\newcommand{\remark}[1]{\medskip \noindent{\bf Remark:} {#1} \par \medskip}
\renewcommand{\and}{{\qquad {\rm and} \qquad}}

\newcommand{\virg}{{\qquad , \qquad}}

\newcommand{\C}{{\mathbf C\,}}
 
\newcommand{\tr}{{\,\rm tr}\:}

\newcommand{\td}[1]{{\tilde{#1}}}

\renewcommand{\l}{\lambda}
\renewcommand{\L}{\Lambda}

\newcommand{\e}{\epsilon}
\newcommand{\ee}[1]{{{\rm e}^{#1}}}

\renewcommand{\d}{{{\partial}}}

\newcommand{\Pint}{{\int\kern -1.em -\kern-.25em}}

\renewcommand{\Re}{{\mathrm{Re}}}
\renewcommand{\Im}{{\mathrm{Im}}}

\newcommand\Res {\mathop{\rm Res}}

\newcommand\xx {{\cal X}}
\newcommand\yy {{\cal Y}}
\newcommand\xxs {{\underline{\xx}}}
\newcommand\yys {{\underline{\yy}}}
\renewcommand\ss {\sigma}
\renewcommand\tt {\td{\sigma}}
\newcommand\wei {{\phi}}

\begin{document}
\sloppy


\pagestyle{empty}
\begin{flushright}
\hfill SPhT-T03/106\
\end{flushright}

\addtolength{\baselineskip}{0.20\baselineskip}
\begin{center}
\vspace{26pt}
{\large \bf {Large $N$ expansion of the 2-matrix model, multicut
case}}
\newline
\vspace{26pt}

{\sl B.\ Eynard}\hspace*{0.05cm}\footnote{ E-mail:
eynard@spht.saclay.cea.fr
}\\
\vspace{6pt}
Service de Physique Th\'{e}orique de Saclay,\\
F-91191 Gif-sur-Yvette Cedex, France.\\
\end{center}

\vspace{20pt}
\begin{center}
{\bf Abstract}
\end{center}
%

\begin{center}
We present a method, based on loop equations, to compute
recursively,
 all the terms in the large $N$ topological expansion of the free
energy for
the 2-hermitian matrix model, in the case where the support of the density of
eigenvalues is not connected.

We illustrate the method by computing the
free energy of a statistical physics model on a discretized
torus.
\end{center}



\newpage
\pagestyle{plain}
\setcounter{page}{1}


\newsection{Introduction}

Random matrix models \cite{Mehta, courseynard, ZJDFG, Guhr, BI, Moerbeke:2000}
have a wide range of applications in mathematics and physics where they
constitute a major field of activity.
They are involved in condensed matter physics (quantum chaos \cite{Guhr,
QChaosHouches}, localization, crystal growths \cite{spohn},...etc),
 statistical physics \cite{ZJDFG, Matrixsurf, BIPZ, grossGQ2D} (on a 2d
fluctuating surface, also called 2d euclidean quantum gravity, linked to
conformal field theory),
 high energy physics (string theory \cite{Dijgrafvafa, BECK}, quantum gravity \cite{ZJDFG,
grossGQ2D, GinspargGQ2D}, QCD \cite{Verbaarshot},...),
 and they are very important in mathematics too: (they seem to be linked to the
Riemann conjecture \cite{Mehta, Odl}), they are important in combinatorics,
 and provide a wide class of integrable systems \cite{Wiegmann, BI, HTW, McLaughlin}.

In the 80's, random matrix models were introduced as a toy model for
zero--dimensional string theory and quantum gravity
\cite{Matrixsurf, ZJDFG, BIPZ}.

The free energy of matrix model is conjectured\footnote{There is at the present time no rigorous proof of
the existence of the topological expansion; The Riemann-Hilbert
approach seems to be the best way to prove it as in \cite{EML}.
The Riemann-Hilbert problem for the 2-matrix model has been formulated
\cite{BEHRH, BEHAMS, BEHansatz, Kapaev}, and
seems to be on the verge of being solved \cite{BEHansatz}.}
 to have a $1/N^2$ expansion \cite{thoft, ZJDFG, ACM, ACKM} called topological expansion ($N$ is the size of the matrix):
\beq
F = \sum_{h=0}^\infty N^{-2h} F^{(h)}
\eeq
That expansion is the main motivation for applications to 2-dimensional
quantum gravity \cite{ZJDFG, courseynard},
 because each $F^{(h)}$ is the partition function of a statistical physics
model on a genus $h$ surface.

The authors of \cite{ACM} invented an efficient method to compute recursively
all the $F^{(h)}$'s for the 1-matrix model, and they improved it in \cite{ACKM}.

Their method was generalized in \cite{eynm2m} for the 2-matrix model in the so-called
1-cut case.

Here, we extend the result of \cite{eynm2m}, to multicut cases.

Assuming that the $1/N^2$ expansion exists, the aim of the present work is to give a method
to compute recursively the terms of the expansion, similar to that of \cite{eynm2m}.

\bigskip

The 2-matrix model \cite{twomatrixHeisenberg, DKK} was first introduced as a model for two-dimensional gravity,
with matter, and in particular with an Ising field \cite{Kazakov,
KazakovIsing}.
The diagrammatic expansion of the 2-matrix model's partition function is known
to generate 2-dimensional statistical physics models on a random discrete
surface \cite{ZJDFG, Matrixsurf, Kazakov}:
\beq
N^2 F=-\ln{Z} = \sum_{\rm surfaces}\,\,  \sum_{\rm matter}  \ee{-{\rm Action}}
\eeq
where the Action is the matter action (like Ising's nearest neighboor spin
coupling) plus the gravity action (total curvature and cosmological constant)
\cite{ZJDFG}.
The cosmological constant couples to the area of the surface, and $N$ (the size
of the matrix) couples to the total curvature, i.e. the genus of the surface.
The large $N$ expansion thus generates a genus expansion:
\beq
F=\sum_{h=0}^\infty N^{-2h} F^{(h)}
\eeq
where $F^{(h)}$ is the partitrion function of the statistical physics model on
a random surface of fixed genus $h$.
\beq\label{Fhsurfaces}
F^{(h)} = \sum_{{\rm genus}\, h\, {\rm surfaces}}\,\,  \sum_{\rm matter}
\ee{-{\rm Action}}
\eeq
The leading term $F^{(0)}$ computed by \cite{Bertolafreeenergy} (along the method invented by
\cite{matytsin} and rigorously established by \cite{guionnet}) is the planar contribution.
Our goal in this article is to compute $F^{(1)}$
and present an algorithmic method for computing $F^{(h)}$ for $h\geq 1$.
We generalize the method of \cite{eynm2m}.

\subsection{Outline of the article}

\begin{itemize}
\item In section \ref{section2matintro} we introduce the definitions and
notations, in particular we define the 1-loop functions and 2-loop functions,
 loop-insertion operators, and we write the ``Master loop equation''.

\item In section \ref{sectionleadingorder}, we observe that, to leading
order, the master loop equation is an algebraic equation of genus zero,
and we study the geometry and the sheet-structure of the underlying algebraic
curve.

\item In section \ref{sectionsubleading} we include the previously neglected
$1/N^2$ term in the loop equation,
and we compute the 1-loop function $Y(x)$ to next to leading order.
Then we derive the next to leading order free energy $F^{(1)}$ by integrating
$Y^{(1)}(x)$.
We also discuss how to compute higher order terms.

\item In section \ref{sectiongenus1} we complete the calculation for the case where
the algebraic curve has genus one.

\item section \ref{conclusion} is the conclusion.

\end{itemize}

\newsection{The 2-matrix model}
\label{section2matintro}
Let $N$ be an integer, $V_1$ and $V_2$ two polynomials of degrees $d_1+1$ and
$d_2+1$:
\beq
V_1(x) = g_0 + \sum_{k=1}^{d_1+1} {g_k\over k} x^k
\virg
V_2(y) = \td{g}_0 + \sum_{k=1}^{d_2+1} {\td{g}_k\over k} y^k
\eeq
Then let $g$ be an integer (called the genus) choosen between $0\leq g \leq d_1
d_2-1$,
and let $\vec\e$ be a $g$ dimensional vector:
\beq
\vec\e = (\e_1,\e_2,\dots,\e_g)^t
\eeq
We define
\beq
\e_{g+1} = 1-\sum_{i=1}^g \e_i
\eeq

All those numbers given, we define the partition function $Z$ and the free
energy $F$ by the following matrix integral:
\beq\label{Zdef}
Z = \ee{-N^2 F} = \int d M_1 d M_2 \,\, \ee{-N \tr [ V_1(M_1) + V_2(M_2) - M_1
M_2 ]}
\eeq
where the integral is over pairs of hermitian matrices $M_1$ and $M_2$
restricted by the following condition:\par
- the large $N$ limit of the density of eigenvalues of $M_1$ has a support made
of $g+1$ disconnected intervals.\par
- the filling fraction (integral of the density) in each interval is $\e_i$.\par

\medskip
We assume that the free energy $F$ admits a topological $1/N^2$ expansion:
\beq
F = F^{(0)} + {1\over N^2} F^{(1)} + \dots + {1\over N^{2h}} F^{(h)} +\dots
\eeq
That asumption plays a key role in many areas of physics, in particular
quantum gravity or string theory \cite{Dijgrafvafa, ZJDFG}, and is believe to hold
for a wide class of potentials. However, the existence of the $1/N^2$ expansion for the 2-matrix model
has never been proven rigorously (for the one matrix model, it has been established by \cite{EML}).

The goal of this article is to develop a method to compute $F^{(h)}$ by
recursion on $h$. In particular, we will explicitely compute $F^{(1)}$.
Notice that this was already done in \cite{eynm2m} in the case $g=0$.
Notice that $F^{(0)}$ was computed by \cite{Bertolafreeenergy}.

\bigskip

\noindent{\bf Remark:}
The model can be extended to normal matrices with support on complex paths, i.e. the eigenvalues are
located along some line in the complex plane, not necessarily the real axis.
In that case, the potentiasl need not have even degrees and positive leading coefficient, 
the potentials can be arbitrary complex polynomials, and the complex paths have to be chosen so that
the partition function \eq{Zdef} makes sense.

\subsection{Definition: resolvents}

We define:
\beq
T_{k,l} := {1\over N} \left< \tr M_1^k M_2^l \right>
\eeq

The resolvents are formaly\footnote{Formaly means the following: the sums in the
RHS are not necessarily convergent.
$W_1(x)$ is merely a convenient notation to deal with all $T_{k,0}$ at once.}
 defined by:
\beq\label{defresolvents}
W_1(x) := \sum_{k=0}^\infty {T_{k,0}\over x^{k+1}}
\virg
W_2(y) := \sum_{l=0}^\infty {T_{0,l}\over y^{l+1}}
\eeq
in other words:
\beq
W_1(x) = {1\over N}\left< \tr {1\over x-M_1} \right>
\virg
W_2(y) = {1\over N}\left< \tr {1\over y-M_2} \right>
\eeq
We also define:
\beq\label{defXY}
Y(x) := V'_1(x)-W_1(x)
\virg
X(y) := V'_2(y)-W_2(y)
\eeq

We assume that all the $T_{k,l}$ have a $1/N^2$ expansion, and we can write
(formaly):
\beq
Y(x)= Y^{(0)}(x) + {1\over N^2}Y^{(1)}(x) +\dots + {1\over N^{2h}}Y^{(h)}(x) +
\dots
\eeq
\beq
X(y)= X^{(0)}(y) + {1\over N^2}X^{(1)}(y) +\dots + {1\over N^{2h}}X^{(h)}(y) +
\dots
\eeq

We will recall below that the leading terms $Y^{(0)}(x)$ and $X^{(0)}(y)$ are
solutions of algebraic equations.
Then we will explain how to compute the first subleading term $Y^{(1)}(x)$.
We will show that we can compute all $Y^{(h)}$ by recursion on $h$.

\subsection{Other 1-loop functions}

We define the following formal functions:
\beq\label{defWxy}
W(x,y) := \sum_{k=0}^\infty \sum_{l=0}^{\infty} {T_{k,l}\over x^{k+1}y^{l+1}}
= {1\over N}\left< \tr {1\over x-M_1} {1\over y-M_2}\right>
\eeq
\beq\label{defUxy}
U(x,y) := \sum_{k=0}^\infty \sum_{j=0}^{d_2} \sum_{l=0}^{j-1} g^*_{j+1}
{y^{j-1-l}\over x^{k+1}} T_{k,l}
= {1\over N}\left< \tr {1\over x-M_1} {V'_2(y)-V'_2(M_2)\over y-M_2}\right>
\eeq
\bea\label{defPxy}
P(x,y) & := & \sum_{i=0}^{d_1} \sum_{j=0}^{d_2} \sum_{k=0}^{i-1} \sum_{l=0}^{j-1}
g_{i+1}g^*_{j+1} x^{i-1-k} y^{j-1-l} T_{k,l} \cr
& = & {1\over N}\left< \tr {V'_1(x)-V'_1(M_1)\over x-M_1}
{V'_2(y)-V'_2(M_2)\over
y-M_2}\right>
\eea
Notice that $P(x,y)$ is a polynomial in $x$ and $y$ of
degree $d_1-1,d_2-1$.

All those functions have a $1/N^2$ expansion.

\subsection{Loop insertion operators}

We define formaly the loop insertion operators:
\beq\label{defdV}
{\d \over \d V_1(x)} := {1\over x} {\d \over \d g_0} + \sum_{k=1}^\infty {k\over x^{k+1}} {\d \over \d g_k}
\virg
{\d \over \d V_2(y)} := {1\over y} {\d \over \d \td{g}_0} + \sum_{k=1}^\infty {k\over y^{k+1}} {\d \over \d g^*_k}
\eeq
These formal definitions actualy mean that for any observable $f$:
\beq
{\d f\over \d g_k} = \Res x^k {\d f\over \d V_1(x)} dx
\eeq

In particular with the free energy, we read from the partition function:
\beq\label{WdFdV}
W_1(x) = {\d \over \d V_1(x)} F
\virg
W_2(y) = {\d \over \d V_2(y)} F
\eeq

\subsection{2-loop functions}

We define the following functions:

\beq\label{defOmega}
\Omega(x;x') := {\d \over \d V_1(x')} W_1(x)
= -\left< \tr {1\over x-M_1}\tr {1\over x'-M_1} \right>_{\rm conn}
\eeq
\beq
\td\Omega(y;x') := {\d \over \d V_1(x')} W_2(y)
= -\left< \tr {1\over x'-M_1}\tr {1\over y-M_2} \right>_{\rm conn}
\eeq
\beq
U(x,y;x') := -{\d \over \d V_1(x')} U(x,y)
= \left< \tr {1\over x-M_1}{V'_2(y)-V'_2(M_2)\over y-M_2} \tr {1\over x'-M_1}
\right>_{\rm conn}
\eeq

\subsection{Master loop equation}

It is shown in \cite{eynm2m,eynard,eynardchain} that we have the following system of equations,
called the "master loop equations" \cite{staudacher}:
\beq\label{loopequation}
\encadremath{
E(x,Y(x)) = {1\over N^2} U(x,Y(x);x)
}\eeq
\beq\label{loopeqU}
U(x,y) = x-V'_2(y) +{E(x,y)\over y-Y(x)} - {1\over N^2} {U(x,y;x)\over y-Y(x)}
\eeq
\beq\label{loopeqUU}
U(x,y;x') = -{\d \over \d V_1(x')} U(x,y)
\eeq
where $E(x,y)$ is a polynomial in $x$ (degree $d_1+1$) and $y$ (degree $d_2+1$):
\beq\label{defE}
\encadremath{
E(x,y) := (V'_1(x)-y)(V'_2(y)-x) - P(x,y) +1
}\eeq
and where $P(x,y)$ was defined in \eq{defPxy}.

We will solve that system of equations, order by order in $1/N^2$:
\beq
Y(x)= Y^{(0)}(x) + {1\over N^2}Y^{(1)}(x) +\dots + {1\over N^{2h}}Y^{(h)}(x) +
\dots
\eeq

\bigskip
Then, the large $N$ expansion of the free energy is obtained from \eq{WdFdV}:
\beq
{\d \over \d V_1(x)} F^{(h)} = \delta_{h,0} V'_1(x) - Y^{(h)}(x)
\eeq

\newsection{leading order: algebraic geometry}
\label{sectionleadingorder}

To leading order in $1/N^2$, the master loop equation
is an algebraic equation for $Y^{(0)}(x)$:
\beq
E^{(0)}(x,Y^{(0)}(x)) = 0
\eeq
where
\beq
E^{(0)}(x,y) = (V'_1(x)-y)(V'_2(y)-x) - P^{(0)}(x,y) +1
\eeq
and $P^{(0)}(x,y)$ is a polynomial of degree $(d_1-1,d_2-1)$ whose coefficient
of
$x^{d_1-1}y^{d_2-1}$ is known and is equal to $g_{d_1+1}g^*_{d_2+1}$ (from \eq{defPxy}).

Before going to order $1/N^2$ and higher, we need to introduce some concepts of algebraic geometry,
and study the geometry of the above algebraic equation.

\subsection{Determination of $P^{(0)}(x,y)$}

So far, we don't know the other $d_1 d_2 -1$ unknown coefficients of
$P^{(0)}(x,y)$.
They are determined by the following requirements:\par
- $E^{(0)}(x,y)=0$ is a genus $g$ algebraic curve.
If $g$ is less than the maximal genus\footnote{The maximal genus is $d_1d_2-1$.
It can be computed by various methods. In particular, see \cite{KazMar} for the
Newton's polygons method} we get $d_1 d_2-1-g$ constraints on the coefficients
of $P^{(0)}(x,y)$.\par
- The contour integrals of $Y^{(0)}(x)dx$ along $A$-cycles are:
\beq\label{acycles}
{1\over 2i\pi} \oint_{{\cal A}_i} Y^{(0)}(x) dx = \e_i\quad , i=1,\dots,g+1
\eeq
where ${\cal A}_i$,${\cal B}_i$, $i=1,\dots,g$ is a canonical basis of irreducible cycles
on the algebraic curve.
We get $g$ independent equations for the coefficients of $P^{(0)}(x,y)$.\par

Therefore we can determine the polynomial $P^{(0)}(x,y)$ (and thus the
polynomial  $E^{(0)}(x,y)$) completely.

\medskip

A physical picture is that the support of the large $N$ average density of eigenvalues of the matrix $M_1$
is made of $g+1$ intervals $[a_i,b_i]$, $i=1,\dots,g+1$, and for each $i$, ${\cal A}_i$ is a contour which
encloses $[a_i,b_i]$ in the trigonometric direction, and which does not enclose the other $a_j$ or $b_j$ with
$j\neq i$. \eq{acycles} means that the interval $[a_i,b_i]$ contains a proportion $\e_i$ of the total number of eigenvalues,
this why $\e_i$ is called a filling fraction.

\bigskip

In practice, we don't have a closed expression for $P^{(0)}(x,y)$ as a function of the coefficients of
$V_1$ and $V_2$ and the filling fractions $\e_i$.
The converse is easier: given a genus $g$ algebraic curve, we can determine $V_1$, $V_2$ and the $\e_i$'S.

\subsection{Algebraic geometry}

Before proceeding, we need to study the geometry of our algebraic curve.

Let us call ${\cal E}$ the algebraic curve, and we consider that an abstract point $p\in{\cal E}$
is a pair of complex numbers $p=(x,y)$ such that $E^{(0)}(x,y)=0$.
Thus, $x=\xx(p)$ and $y=\yy(p)$ are complex-valued functions on the curve.

Here, we sumarize some well known properties of Riemann surfaces and theta-functions.
We refer the interested reader to textbooks \cite{Farkas, Fay} for proofs and complements.

\subsection{Sheet structure}

The function $Y^{(0)}(x)$ (resp. $X^{(0)}(y)$) is multivalued,
it takes $d_2+1$ (resp. $d_1+1$) values, which we note:
\beq
(Y_0(x),Y_1(x),\dots,Y_{d_2}(x))
\qquad ({\rm resp}.\,\,
(X_0(y),X_1(y),\dots,X_{d_1}(y)) \,)
\eeq
The ${\rm zero}^{\rm th}$ sheet is called the physical sheet, it is the one such that (from \eq{defXY}):
\beq\label{behaviourphyssheet}
Y_0(x) \mathop\sim_{x\to\infty} V'_1(x) - {1\over x} + O(1/x^2)
\qquad ({\rm resp}.\,\,
X_0(y) \mathop\sim_{y\to\infty} V'_2(y) - {1\over y} + O(1/y^2)
\,)
\eeq

An $x$-sheet (resp. $y$-sheet) is a domain of ${\cal E}$ on which the function $\xx$ (resp. $\yy$)
is a bijection with $\C\cup\{+\infty\}$.
The curve ${\cal E}$ is thus decomposed into $d_2+1$ $x$-sheets (resp. $d_1+1$ $y$-sheets).
The decomposition is not unique, and a cannonical possible decomposition will be given below.

For each $x$ (resp. $y$), there are exactly $d_2+1$ (resp. $d_1+1$) points on ${\cal E}$, one in each sheet, such that:
\bea
&& \xx(p)=x \quad\leftrightarrow\, p=p_k(x) \quad k=0,\dots,d_2 \cr
&\qquad ({\rm resp}.\,\,
&  \yy(p)=y \quad\leftrightarrow\, p=\td{p}_k(y) \quad k=0,\dots,d_1
\,)
\eea
And thus:
\beq
Y_k(x)=\yy(p_k(x))
\qquad ({\rm resp}.\,\,
X_k(y)=\xx(\td{p}_k(y))
\,)
\eeq

\subsection{Points at $\infty$, poles of $\xx$ and $\yy$}

In particular, in each $x$-sheet, there is a point $p$ such that $\xx(p)=\infty$.
It can be seen from \eq{behaviourphyssheet}, that there are exactly two such points on ${\cal E}$.
We define $p_\pm$ such that:
\beq
p\to p_+
\leftrightarrow
\left\{\begin{array}{l}
\xx(p) \to\infty \cr
\yy(p)\to\infty \cr
\yy(p)\sim V'_1(\xx(p))
\end{array}\right.
\qquad
{\rm and}
\qquad
p\to p_-
\leftrightarrow
\left\{\begin{array}{l}
\xx(p) \to\infty \cr
\yy(p)\to\infty \cr
\xx(p)\sim V'_2(\yy(p))
\end{array}\right.
\eeq
$p_+$ (resp. $p_-$) is in the $x$-physical sheet (resp. $y$-physical sheet), while $p_-$ (resp. $p_+$)
is at the intersection of the other $d_2$ $x$-sheets (resp. $d_1$ $y$-sheets).

\subsection{Endpoints and cuts}

The $x$-endpoints (resp. $y$-endpoints) correspond to singularities of $Y(x)$ (resp. $X(y)$),
i.e. they are such that $dY/dx=\infty$ (resp. $dX/dy=\infty$), i.e. they are the zeroes
of $d\xx(p)$ (resp. $d\yy(p)$). There are $d_2+1+2g$ (resp. $d_1+1+2g$) such endpoints:
\beq
d\xx(p) = 0 \, \leftrightarrow \,
p\in\{e_1,e_2,\dots,e_{d_2+1+2g}\}
\,\, , \,\,
d\yy(p) = 0 \, \leftrightarrow \,
p\in\{\td{e}_1,\dots,\td{e}_{d_1+1+2g}\}
\eeq
the endpoints are such that $\exists k\neq l$, $p_k(x)=p_l(x)$ (resp. $\td{p}_k(y)=\td{p}_l(y)$), i.e.
they are at the intersection of two sheets.

\subsubsection{Critical points}

In a generic situation, all the endpoints are distinct.
If $V_1$, $V_2$ and $\e_i$ are chosen so that some endpoints coincide, we say that we are at a critical point.
Imagine that $e$ is such a multiple endpoint, near which $d\xx$ has a zero of degree $q-1$ and $d\yy$ has a zero
of degree $p-1$, then:
\beq
Y(x) -\yy(e) \sim O\left( (x-\xx(e))^{p/q}\right)
\eeq
which is a typical critical behaviour of a $(p,q)$ conformal minimal model.

From now on, we assume that we are in a generic situation, i.e. all the endpoints are distinct.

\subsubsection{Cuts}

The cuts are the contours which border the sheets.
Like the sheets, they are not uniquely defined, there is some arbitrariness.

A canonical choice for the cuts is the following:
the cuts are the sets of $p\in {\cal E}$ such that
\beq
\exists q\neq p ,\,\,\, \xx(p)=\xx(q) \,\, {\rm and} \,\, \Re \int_{p}^{q} \yy(u)d\xx(u) = 0
\eeq

\subsection{Irreducible cycles}

We have already introduced a basis of irreducible cycles,
${\cal A}_i$, ${\cal B}_i$,$i=1,\dots,g$, such that:
\beq
{\cal A}_i \bigcap {\cal B}_j = \delta_{i,j}
\eeq
Moreover, we assume that the $A$-cycles are cuts, and that the $A$ and $B$ cycles do not intersect a line $L$
which joins $p_+$ and $p_-$.

We have:
\beq\label{Acycleei}
{1\over 2i\pi}\oint_{{{\cal A}_i}} \yy(p) d\xx(p) = \epsilon_i
\eeq
and we define:
\beq\label{Bcycleei}
\Gamma_j
:= \oint_{{{\cal B}_j}} \yy(p) d\xx(p)
\eeq
We then define the period-matrix $\tau$ by:
\beq
\tau_{i,j} := {1\over 2i\pi}{\d \Gamma_i\over \d \epsilon_j}
\eeq

\bigskip
{Remark:}\par
It is proven in the appendix \ref{appendixcalcul} that:
\beq
\Gamma_j
= {\d F^{(0)}\over \d \epsilon_j}
\virg
\tau_{i,j} = {1\over 2i\pi} {\d^2 F^{(0)}\over \d \epsilon_i \epsilon_j}
\eeq
where $F^{(0)}$ is computed from \eq{WdFdV}.
Notice that $\tau$ is symmetric.

\subsection{Holomorphic differentials}

We define the following differential one-forms:
\beq\label{defdui}
d{u_i}(p) := {1\over 2i\pi} {\d \over \d\epsilon_i} \left(\yy(p) d\xx(p)\right)
\eeq
they are holomorphic.
Indeed, the pole of $\yy(p)d\xx(p)=d V_1(\xx(p)) - W_1(\xx(p)) d\xx(p)$ at $p=p_+$
is independent of $\e_i$ (because $V_1$ is independent of $\e_i$ and Res$W_1(x)dx=1$ is independent of $\e_i$),
therefore $d{u_i}(p)$ has no pole at $p_+$.
By the same argument,  $d{u_i}(p)$ has no pole at $p_-$, and  $d{u_i}(p)$ is holomorphic.
Moreover we have (from \eq{Acycleei} and \eq{Bcycleei}):
\beq
\oint_{{{\cal A}_i}} d{u_j}(p) = \delta_{i,j}
\virg
\oint_{{{\cal B}_i}} d{u_j}(p) = \tau_{i,j}
\eeq

\bigskip
{Remark:}\par
Anticipating a little bit, it follows from \eq{Bcycleei}, \eq{defB} and \eq{propertyB},
 that:
\beq\label{duidGammadV}
d{u_i}(p) = -{1\over 2i\pi}{\d \Gamma_i\over \d V_1(\xx(p))} d{\xx(p)}
\eeq

\subsection{Abelian differential of the third kind}

On the Riemann surface ${\cal E}$, there exists a unique abelian differential of the third kind $dS$,
with two simple poles at $p=p_\pm$, such that:
\beq\label{defdS}
\Res_{p_+} dS = -1 = -\Res_{p_-} dS
\quad {\rm and} \quad
\forall i \,\,\,\oint_{{{\cal A}_i}} dS = 0
\eeq

We choose an arbitrary point $p_0\in {\cal E}$, which does not belong to any cut or any irreducible cycle,
and we choose a line $L$ joining $p_+$ to $p_-$, which does not intersect any cycle and does not contain $p_0$.
Then we define the following functions on ${\cal E}\backslash (\cup_i {\cal A}_i \, \cup_i {\cal B}_i \, \cup L) $:
\beq\label{defSL}
S(p):= \int_{p_0}^p dS
\virg
\L(p) := \exp{S(p)}
\eeq
where the line of integration does not intersect any cycle neither the line $L$
(notice that the integral around $L$ vanishes because it encloses $p_+$ and $p_-$ which have opposite residues).
We have:
\beq
\L(p) = {E(p,p_-)\over E(p,p_+)}
\eeq

$S(p)$ has logarithmic singularities near $p_+$ and $p_-$, and is discontinuous along $L$, the discontinuity is:
\beq
\delta S(p) = 2i\pi \qquad p\in L
\eeq

$S(p)$ is continuous along the $B$-cycles, and discontinuous along the $A$-cycles, the discontinuity is:
\beq
\delta S(p) = \eta_i \qquad p\in {\cal A}_i
\virg
\eta_i :=\oint_{{\cal B}_i} dS = u_i(p_+)-u_i(p_-)
\eeq

$\L(p)$ has no discontinuity along $L$, it has a simple pole at $p_+$,
and a simple zero at $p_-$, therefore the following quantities are well defined:
\beq\label{defgamma}
\gamma := \mathop{\rm lim}_{p\to p_+} \xx(p)/\L(p)
\virg
\td\gamma := \mathop{\rm lim}_{p\to p_-} \yy(p)\L(p)
\eeq

\bigskip
Remark:
By an appropriate choice of $p_0$, it should be possible to have $\gamma=\td\gamma$, however,
we will not make that asumption.

\section{2-loop functions and the Bargmann kernel}

Consider the 2-loop function defined in \eq{defOmega}:
\beq
\Omega(x;x')= {\d W_1(x)\over \d V_1(x')} = {1\over (x-x')^2} -
{\d Y(x)\over \d V_1(x')} = -\left< \tr {1\over x-M_1}\tr {1\over
x'-M_1} \right>_{\rm conn}
\eeq
and define the bilinear differential (where $x=\xx(p)$ and $x'=\xx(p')$):
\beq\label{defB}
B(p,p') := {\d Y(x)\over \d V_1(x')}\, dx\, dx'
\eeq

$\Omega(x;x')$ has the following properties:
\begin{itemize}
\item $\Omega(x;x')=\Omega(x';x)$ is symmetric.

\item since $Y(x)$ has square root singularities near the endpoints $e_k$,
$\Omega(x;x')$ has inverse square
root singularities near the endpoints (i.e. simple poles in $p\to e_k$).
Therefore $B(p,p')$ has no pole in $p=e_k$.

\item since $W_1(x)$ behaves like ${1\over x} + O(1/x^2)$ in the physical sheet,
i.e. when $x\to p_+$, we must have:
$\Omega(x;x') \sim O(1/x^2)$. In particular
$B(p,p')$ is finite when $p\to p_+$.

\item since $Y(x)$ behaves like $V'_2(Y(x)) - {1\over Y(x)} +O(1/Y^2(x)) \sim x$
when $x\to p_-$, we must have that
$B(p,p')$ is finite when $p\to p_-$.

\item $\Omega(x;x')$ has no pole at $x=x'$ in the same sheet (i.e. when $p=p'$).
This implies that $B(p,p')\sim (\xx(p)-\xx(p'))^{-2} dx(p) dx(p')$ when $p\to p'$.

\item since $Y(x)$ satisfies \eq{acycles} and $\epsilon$ is independent on
$V_1$, we must have:
\beq\label{acycles1Omega}
\forall i \,\,\, \int_{{\cal A}_i} \Omega(x;x') d x =0
\eeq

\end{itemize}

This allows to determine $\Omega(x;x')$.
Indeed, $B(p,p')$ is a meromorphic bilinear differential on ${\cal E}$,
with only one normalized double pole at $p=p'$, and normalized $A$-cycles,
therefore $B(p,p')$ is the Bargmann kernel, i.e. the unique meromorphic bilinear differential with such properties.

It can be written (see appendix \ref{thetagenusall}):
\beq
B(p,p') = \d_i\d_j \ln\theta{(\vec{u}(p)-\vec{u}(p')-\vec{z})}\,\,\, du_i(p) du_j(p')
\eeq

It has the property that:
\beq\label{propertyB}
\oint_{p\in{\cal B}_i} B(p,p') = 2i\pi du_i(p')
\eeq

Notice that in \eq{defB}, the derivative is taken at fixed $x=\xx(p)$.

\section{$1/N^2$ Expansion}
\label{sectionsubleading}

We are now interested in the $1/N^2$ expansion of the free energy and loop functions:
\beq
F = F^{(0)} + {1\over N^2} F^{(1)} +\dots
\virg
Y(x) = Y^{(0)}(x) + {1\over N^2} Y^{(1)}(x) +\dots
\eeq
where
\beq\label{YdVdFns}
{\d \over \d V_1(x)} F^{(h)} = \delta_{h,0} V'_1(x) - Y^{(h)}(x)
\eeq
So far, we have explained how to compute $Y^{(0)}(x)$.
Once $Y^{(0)}(x)$ is known, $F^{(0)}$ can in principle be computed from \eq{YdVdFns},
this has been done in \cite{Bertolafreeenergy}.

Our goal is to compute $Y^{(1)}(x)$, $F^{(1)}$,
and then define a recursive procedure to compute $Y^{(h)}(x)$ and $F^{(h)}$ for all $h>1$.

\subsection{$1/N^2$ term}

First we expand the polynomial $E(x,y)$:
\beq
E(x,y) = E^{(0)}(x,y) + {1\over N^2} E^{(1)}(x,y) + \dots
\eeq
where
$E^{(1)}(x,y)=-P^{(1)}(x,y)$ is a polynomial of degree $(d_1-1,d_2-1)$ whose
coefficient of $x^{d_1-1}y^{d_2-1}$ vanishes.
And we write similar expansions for all other loop functions, in particular $U(x,y)$
and $U(x,y;x')$.
\beq
U(x,y) = U^{(0)}(x,y) + {1\over N^2} U^{(1)}(x,y) + \dots
\eeq
\beq
U(x,y;x') = U^{(0)}(x,y;x') + {1\over N^2} U^{(1)}(x,y;x') + \dots
\eeq

Then we expand \eq{loopequation} to order $1/N^2$:
\beq
E^{(1)}(x,Y^{(0)}(x)) + Y^{(1)}(x) E_y^{(0)}(x,Y^{(0)}(x)) =
U^{(0)}(x,Y^{(0)}(x);x)
\eeq
i.e.:
\beq\label{loopeqY1}
Y^{(1)}(x)  = {P^{(1)}(x,Y^{(0)}(x)) + U^{(0)}(x,Y^{(0)}(x);x)  \over
E_y^{(0)}(x,Y^{(0)}(x))}
\eeq
So far, the polynomial $P^{(1)}(x,y)$ is unknown, i.e. we have $d_1 d_2-1$ unknown coefficients.

We expect that order by order in the $1/N^2$ expansion, the resolvent $W_1(x)=V'_1(x)-Y(x)$
has no singularities appart from the endpoints, so we require that
$Y^{(1)}$ has singularities only at the endpoints.

The condition that $Y^{(1)}$ has singularities only at the endpoints, implies that in \eq{loopeqY1},
the poles at the zeroes of $E_y^{(0)}(x,Y^{(0)}(x))$ which are not endpoints should cancel.
Since there are $d_1 d_2 -1 $ such points, we can determine $P^{(1)}(x,y)$,
and thus we can determine $Y^{(1)}(x)$.
In other words, $P^{(1)}$ is determined by the condition that
$Y^{(1)}$ has singularities only at the endpoints.

\subsection{The function $U(x,y;x')$ to leading order}

Consider $x$ in the physical sheet, so that $Y^{(0)}(x)=Y_0(x)$.
From \eq{loopeqU} we have:
\beq
U^{(0)}(x,y;x') = -{\d U^{(0)}(x,y)\over \d V_1(x')}
= -{{\d E^{(0)}(x,y)\over \d V_1(x')}\over y-Y^{(0)}(x)}
- {\d Y^{(0)}(x)\over \d V_1(x')} {E^{(0)}(x,y)\over (y-Y^{(0)}(x))^2}
\eeq

Notice that
\beq
E^{(0)}(x,y) = -g^*_{d_2+1} \prod_{k=0}^{d_2} (y-Y_k(x))
\eeq
(indeed, both sides are polynomials in $y$ with the same degree, the same zeroes
and the same leading term).
Therefore:
\beq
{\d E^{(0)}(x,y)\over \d V_1(x')}
= - E^{(0)}(x,y) \sum_{k=0}^{d_2} {\d Y_k(x)\over \d V_1(x')}\,{1\over
y-Y_k(x)}
\eeq
and thus:
\beq
U^{(0)}(x,y;x')
=  {E^{(0)}(x,y)\over y-Y_0(x)}\,\sum_{k=1}^{d_2} {\d Y_k(x)\over \d V_1(x')}\,{1\over y-Y_k(x)}
\eeq
Notice that we have considerably simplified the derivation given in \cite{eynm2m}

In particular, when $y=Y_0(x)$ and $x'=x$, we have:

\beq\label{UxYx}
U^{(0)}(x,Y_0(x);x)  =   E^{(0)}_y(x,Y_0(x))\,\sum_{k=1}^{d_2} {\d Y_k(x)\over \d V_1(x)}\,{1\over Y_0(x)-Y_k(x)}
\eeq

\subsection{$Y^{(1)}$}

Using \eq{loopeqY1} and \eq{UxYx} we have:
\beq\label{loopeqY1U}
\encadremath{
Y^{(1)}(x)  = {P^{(1)}(x,Y^{(0)}(x))  \over E_y^{(0)}(x,Y^{(0)}(x))}
+ \sum_{k=1}^{d_2} {\d Y_k(x)\over \d V_1(x)}\,{1\over Y_0(x)-Y_k(x)}
}\eeq
and $Y^{(1)}$ has poles (of degree up to 5) only at the endpoints.
In other words, $Y^{(1)}(x) dx$ is a one form, with poles only at the endpoints (no pole near $p_+$ and $p_-$).

\subsubsection{Behaviour near the endpoints}

Recall that the endpoints are the zeroes of $d\xx(e)=0$.
If $p$ is near an endpoint $e_k$, there exists a unique (because we have assumed that the potentials are generic)
$p'$ such that $\xx(p')=\xx(p)$ and $p'$ is near $e_k$.

We have
\beq
Y^{(1)}(x(p)) d\xx(p) = {B(p,p')\over d\xx(p')} \,{1\over \yy(p)-\yy(p')} + O(1) \quad {\rm when}\,\, p\to e_k
\eeq
where $B(p,p')$ is the Bargmann kernel.

This can also be written:
\beq
Y^{(1)}(x(p)) d\xx(p) = \Res_{p''\to p'} {B(p,p'')\over (\xx(p)-\xx(p''))(\yy(p)-\yy(p''))} + O(1) \quad {\rm when}\,\, p\to e_k
\eeq

By adding only $O(1)$ quantities, we arrive at:
\beq
Y^{(1)}(x(p)) d\xx(p) = \Res_{p''\to p} {B(p,p'')\over (\xx(p)-\xx(p''))(\yy(p)-\yy(p''))} + O(1) \quad {\rm when}\,\, p\to e_k
\eeq

Since that quantity is symmetric in $x$ and $y$, we have:
\beq
\encadremath{
\displaystyle 
\begin{array}{ll}
Y^{(1)}(\xx(p)) d\xx(p) + X^{(1)}(\yy(p)) d\yy(p)
 = & \displaystyle \Res_{p'\to p}\, {B(p,p')\over (\xx(p)-\xx(p'))(\yy(p)-\yy(p'))} \cr
 & + \displaystyle \sum_{i=1}^g C_i du_i(p)
 \end{array}
}\eeq
where $C_i$ are some constants.
Indeed, the difference between the LHS and RHS has no pole, it is a holomorphic one-form.

\subsubsection{local coordinate near an endpoint}

Consider that $z(p)$ is a local coordinate near an endpoint $e_k$, we have:
\beq
\xx(p) = \xx(e_k) + {z^2\over 2} \xx''(e_k) + {z^3\over 6} \xx'''(e_k)  + {z^4\over 24} \xx^{IV}(e_k) + \dots
\eeq
\beq
\yy(p) = \yy(e_k) + z \yy'(e_k) + {z^2\over 2} \yy''(e_k) + {z^3\over 6} \yy'''(e_k)  + \dots
\eeq
\beq
B(p,p') = \left( {1\over (z-z')^2} + {1\over 6}S(e_k) + \dots \right)\, dz\, dz'
\eeq
where $S(p)$ is the projective connection.

$\xx(p')=\xx(p)$ implies:
\beq
z' = -z (1+ r_k z + r_k^2 z^2 + (2r_k^3+t_k)z^3 + \dots )
\eeq
where
\beq
r_k = {1\over 3} {\xx'''(e_k)\over \xx''(e_k)}
\virg
s_k = {1\over 6} {\xx^{IV}(e_k)\over \xx''(e_k)}
\virg
t_k = {1\over 60} {\xx^{V}(e_k)\over \xx''(e_k)} -r_k s_k
\eeq
That gives:
\bea
Y^{(1)}(\xx(p)){d\xx(p)\over dz}
& = & {1\over 8 \xx''(e_k)
\yy'(e_k)} z^{-4}
  - {\xx'''(e_k) \over 24
\xx''^2(e_k)  \yy'(e_k)} z^{-3} \cr
&& + {{\xx'''(e_k)^2\over \xx''(e_k)^2}
-  {\xx^{IV}(e_k)\over \xx''(e_k)}
 + {\xx'''(e_k)\over \xx''(e_k)}{\yy''(e_k)\over \yy'(e_k)}
- {\yy'''(e_k)\over \yy'(e_k)}
\over 48 \xx''(e_k) \yy'(e_k)} z^{-2} \cr
&& - {S(e_k)\over 12 \xx''(e_k) \yy'(e_k)} z^{-2} + O(1) \cr
\eea
This is in principle sufficient to determine $Y^{(1)}$.


\subsection{Free energy}

\bigskip
Then, we want to find the free energy $F^{(1)}$ such that:
\beq
Y^{(1)}(x) = -{\d F^{(1)}\over \d V_1(x)}
\eeq
In this purpose, we have to compute the derivatives of various
quantities with respect to $V_1(x)$, and in particular, how the theta-function
parametrization changes with the potential $V_1$.

We conjecture:
\beq
F^{(1)} = -{1\over 24} \ln{\prod_i Y'(e_i)}
\eeq
where
\beq
\ln{Y'(e_i)}:= \int_{p=p_+}^{e_i}\int_{p=p_+}^{e_i} \left( B(p,p')- {dy(p)dy(p')\over (p-p')^2}\right)
\eeq

\subsection{Higher orders}

Imagine we already know all quantities up to order $h-1$, and write \eq{loopequation} to order $h$:
\bea
&& \sum_{j=0}^h N^{-2h+2j} E^{(h-j)}\left(x,\sum_{k=0}^j N^{-2k}  Y^{(k)}(x)\right)  \cr
&& = \sum_{j=0}^{h-1} N^{-2h+2j}  U^{(h-j)}\left(x,\sum_{k=0}^j N^{-2k}  Y^{(k)}(x);x\right)
  +O(N^{-2h})
\eea
The only unknown quantities in that equation are:
$Y^{(h)}(x)$ and $E^{(h)}(x,Y^{(0)}(x))$.
The polynomial $E^{(h)}(x,y)$ must be chosen such that $Y^{(h)}(x)$ has no other singularities
than the endpoints, and is completely determined by this consition.
That allows to find $Y^{(h)}(x)$ as well as $E^{(h)}(x,y)$ and $U^{(h)}(x,y)$ to order $h$.

The procedure can be repeated recursively to find $Y^{(h)}(x)$ to any order.


\newsection{Genus 1 case}
\label{sectiongenus1}

Let us recall that the case $g=0$ was done in \cite{eynm2m}.
The case $g=1$ is treated in this section.

We require that $E(x,y)=0$ be a {\bf genus one} algebraic
curve.
Therefore, there must exist an elliptic uniformization.
We choose it of the following form:

\bea
x=\xx(u) & = & \gamma {\prod_{i=0}^{d_2} \theta(u-\ss_i(0)) \over
\theta(u-u_{\infty})\theta(u+u_{\infty})^{d_2} }
{\theta(2u_\infty)^{d_2+1}\over\prod_i \theta(u_\infty-\ss_i(0))}
\cr
y=\yy(u) & = & \td\gamma {\prod_{i=0}^{d_1} \theta(u-\tt_i(0))
\over  \theta(u-u_{\infty})^{d_1}\theta(u+u_{\infty}) }
{\theta(2u_\infty)^{d_1+1}\over\prod_i \theta(u_\infty+\tt_i(0))}
\eea
and we denote $\tau$ the modulus.
Here, $\theta$ denotes $\theta_1$, i.e. the prime form for genus 1.
A definition of the $\theta$-function and its properties can be found in
appendix \ref{thetagenus1}.

We must have:
\beq\label{conditionalphabeta}
\sum_{i} \ss_i(0) = (d_2-1) u_{\infty}
\virg
\sum_{i} \tt_i(0) = (d_1-1) u_{\infty}
\eeq

All this means that for every $(x,y)$ which satisfy $E(x,y)=0$,
 there exists at least one $u$ (in the fundamental paralellogram of sides $1,\tau$)
such that $x=\xx(u)$ and $y=\yy(u)$.

An alternative parametrization is:
\beq
\xx(u) = \gamma{\theta(2u_\infty)\over \theta'(0)} (Z(u-u_\infty)-Z(u+u_\infty))
+ A_0 + \sum_{k=2}^{d_2} {A_k\over
k-1!}\wei^{(k-2)}(u+u_\infty)
\eeq
\beq
\yy(u) = -\td\gamma{\theta(2u_\infty)\over \theta'(0)}
(Z(u+u_\infty)-Z(u-u_\infty)) + \td{A}_0 + \sum_{k=2}^{d_1}
{\td{A}_k\over k-1!}\wei^{(k-2)}(u-u_\infty)
\eeq
where $Z$ is the Zeta-function, i.e. the log-derivative of $\theta_1$, and $\wei$ is the Weierstrass function, i.e.
$\wei=-Z'$.

\medskip

We note the inverse functions:
\beq
x=\xx(s) \leftrightarrow s=\ss(x)
\virg
y=\yy(s) \leftrightarrow s=\tt(y)
\eeq
The functions $\ss(x)$ and $\tt(y)$ are multivalued, we will
discuss their
sheet structure below.
The functions $Y(x)$ and $X(y)$ are:
\beq
Y(x) = \yy(\ss(x)) \virg X(y) = \xx(\tt(y))
\eeq
They are multivalued too, and their sheet structure will be
discussed below.

\subsection{The parameters}

Our parametrization depends on $d_1+d_2+6$ parameters which are:
 $\ss_k(0)$ ($k=1,\dots,d_2$), $\tt_j(0)$ ($j=1,\dots,d_1$), $\gamma$,
$\td\gamma$, $u_{\infty}$ and $\tau$.
The condition \eq{conditionalphabeta} means that only  $d_1+d_2+4$ of them are
independent.

Equations \eq{behaviourphyssheet} read:
\beq\label{gkparam1}
g_k = {1\over 2i\pi} \oint_{u_\infty} ds {\yy(s)\xx'(s)\over
\xx(s)^k} \qquad k=1,\dots,d_1+1
\eeq
\beq\label{tdgkparam1}
\td{g}_k = {1\over 2i\pi} \oint_{-u_\infty} ds
{\xx(s)\yy'(s)\over \yy(s)^k} \qquad k=1,\dots,d_2+1
\eeq
and
\beq
1 = {1\over 2i\pi} \oint_{u_\infty} ds \yy(s)\xx'(s) = {1\over
2i\pi} \oint_{-u_\infty} ds \xx(s)\yy'(s)
\eeq
Where the contour of integrations are small cylces around $\pm u_\infty$.

And \eq{acycles} reads:
\beq
2i\pi\epsilon = \int_0^\tau ds \,\, \yy(s) \xx'(s)
\eeq
We thus have $d_1+d_2+4$ equations,
therefore we can, in principle, determine all the parameters.

In principle, it should be possible to revert these formula, and
compute the
$\ss(0)$'s and $\tt(0)$'s as functions of the coupling constants
$g$ and $\td{g}$.
This can be done at least numerically.

\remark{Another point of view is interesting too:
once \eq{acycles} and \eq{behaviourphyssheet} are taken into account, we have $d_1+d_2+2$
independent parameters. This is precisely the number of coefficients of the
potentials $V_1$ and $V_2$.
We can consider that {\bf the parameters are merely a reparametrization of the 
coefficients of the potentials}, according to \eq{gkparam1} and
\eq{tdgkparam1}.}

\subsection{endpoints and cuts}

The endpoints in the $x$-plane (resp. $y$-plane), i.e. the 
singularities of $Y(x)$ (resp. $X(y)$) are such that:
\beq
\xx'(s)=0 \qquad ({\rm resp.}\, \yy'(s)=0 )
\eeq
There are $d_2+3$ (resp. $d_1+3$) such endpoints:
\beq
\xx'(s) = 0 \,\, \leftrightarrow \,\,
s\in\{e_1,e_2,\dots,e_{d_2+3}\}
\virg
\yy'(s) = 0 \,\, \leftrightarrow \,\,
s\in\{\td{e}_1,\dots,\td{e}_{d_1+3}\}
\eeq

We can write:
\beq
\xx'(s) = -\gamma {\theta'(0)\theta(2u_\infty)^{d_2+2}\over \prod_i 
\theta(u_\infty-e_i)} \,\, 
{\prod_{i=1}^{d_2+3} \theta(s-e_i) \over 
\theta(s-u_\infty)^2 \theta(s+u_\infty)^{d_2+1} }
\eeq
\beq
\yy'(s) = \td\gamma {\theta'(0)\theta(2u_\infty)^{d_1+2}\over \prod_i
\theta(u_\infty+\td{e}_i)} \,\, {\prod_{i=1}^{d_1+3} \theta(s-\td{e}_i) \over
\theta(s-u_\infty)^{d_1+1} \theta(s+u_\infty)^{2} }
\eeq
Since $\xx'(s)$ and $\yy'(s)$ are elliptical functions, we must have:
\beq\label{condei1}
\sum_i e_i = -(d_2-1) u_\infty \virg \sum_i \td{e}_i = (d_1-1) u_\infty
\eeq
and since $\xx'(s)$ and $\yy'(s)$ are the derivatives of elliptical functions,
we must have:
\beq
\sum_i Z(u_\infty-e_i) = (d_2+1) Z(2u_\infty)
\virg
\sum_i Z(u_\infty+\td{e}_i) = (d_1+1) Z(2u_\infty)
\eeq
Notice that \eq{gkparam1} for $k=d_1+1$ and \eq{tdgkparam1} for $k=d_2+1$ imply:
\beq
\td\gamma = - d_1 g_{d_1+1} \gamma^{d_1} {\prod_i \theta(u_\infty+\td{e}_i)
\over \prod_i \theta(u_\infty-\td{e}_i)}
\virg
\gamma = - d_2 \td{g}_{d_2+1} \td\gamma^{d_2} {\prod_i \theta(u_\infty-e_i)
\over \prod_i
\theta(u_\infty+e_i)}
\eeq

\subsection{2-loop functions and the Bargmann kernel}

Consider the 2-loop function defined in \eq{defOmega}:
\beq
\Omega(x;x')= {\d W_1(x)\over \d V_1(x')} = {1\over (x-x')^2} -
{\d Y(x)\over \d V_1(x')} = -\left< \tr {1\over x-M_1}\tr {1\over
x'-M_1} \right>_{\rm conn}
\eeq
$\Omega(x;x')$ has the following properties:
\begin{itemize}
\item $\Omega(x;x')=\Omega(x';x)$ is symmetric.

\item since $Y(x)$ has square root singularities near the endpoints $\xx(e_k)$,
$\Omega(x;x')$ has inverse square
root singularities near the endpoints (i.e. simple poles in $\ss(x)-e_k$).
Therefore $\xx'(s) \Omega(\xx(s);x')$ is finite when $s\to e_k$.

\item since $W_1(x)$ behaves like ${1\over x} + O(1/x^2)$ when $x\to\infty$
(i.e. $\ss(x)\to +u_\infty$), we must have:
$\Omega(x;x') \sim O(1/x^2)$. In particular
$\xx'(s) \Omega(\xx(s);x')$ is finite when $s\to +u_\infty$.

\item since $Y(x)$ behaves like $V'_2(Y(x)) - {1\over Y(x)} +O(1/Y^2(x)) \sim x$
when $\ss(x)\to -u_\infty$, we must have that
$\xx'(s)\Omega(\xx(s);x')$ is finite when $s\to -u_\infty$.

\item $\Omega(x;x')$ has no pole at $x=x'$ when $x$ and $x'$ are in the same
sheet, i.e. when $\ss(x)=\ss(x')$.
This implies that ${\d Y(x)\over \d V_1(x')} \sim {1\over (x-x')^2}$ when
$\ss(x)\to\ss(x')$.

\item since $Y(x)$ satisfies \eq{acycles} and $\epsilon$ is independent on
$V_1$, we must have:
\beq\label{acycles1Omegag1}
\int_0^1 \Omega(\xx(s);x') \d s =0
\eeq

\end{itemize}

This allows to determine $\Omega(x;x')$.
Indeed, the function $\xx'(s)\xx'(u){\d Y(\xx(s))\over \d V_1(\xx(u))}$
is an elliptical function of $s$, with only one double pole at $s=u$.
Therefore (see appendix \ref{thetagenus1}):
\beq
\xx'(s)\xx'(u){\d Y(\xx(s))\over \d V_1(\xx(u))} = \wei(s-u) + C
\eeq
where $\wei$ is the Weierstrass function, and where the constant $C$ must be
equal to zero in order to satisfy \eq{acycles1Omega}.

We recognize the Bargmann kernel:
\beq\label{OmegaBargmann1}
\encadremath{
{\d Y(x)\over \d V_1(x')} = \ss'(x) \ss'(x')
\wei(\ss(x)-\ss(x')) = \d_x \d_{x'} \ln{\theta(\ss(x)-\ss(x'))}
}\eeq

In a similar fashion, we find that the function:
\beq
\td\Omega(y;x') = {\d W_2(y)\over \d V_1(x')} = -\left< \tr
{1\over y-M_2}\tr{1\over x'-M_1}\right>_{\rm conn}
\eeq
is:
\beq\label{tdOmegaBargmann1}
\encadremath{
\td\Omega(y;x') = - {\d X(y)\over \d V_1(x')} = \tt'(y) \ss'(x')
\wei(\tt(y)-\ss(x')) = \d_y \d_{x'} \ln{\theta(\tt(y)-\ss(x'))}
}\eeq

We are now equipped to compute the next to leading order
functions...

\subsection{Computation of $Y^{(1)}$}

We have (see \eq{loopeqY1U}):
\beq\label{YonePunknown}
Y^{(1)}(x)  =
{P^{(1)}(x,Y_0(x))\over E_y(x,Y_0(x))}
+ \sum_{k=1}^{d_2} {\wei(\ss_k(x)-\ss_0(x))\over \xx'(\ss_k(x))
\xx'(\ss_0(x))(Y_0(x)-Y_k(x))}
\eeq
and we require that $Y^{(1)}(x)$'s only poles (of degree up to 5)
 are the endpoints $\xx(e_k)$.
Note that $\xx'(s) Y^{(1)}(\xx(s))$ has no pole at $s=\pm u_\infty$,
therefore we may write:
\beq\label{Yoneansatzf}
Y^{(1)}(\xx(s)) = {1\over \xx'(s)} \sum_{k=1}^{d_2+3} A_k
\wei''(s-e_k) +
B_k \wei'(s-e_k) + C_k \wei(s-e_k) + D_k Z(s-e_k)
\eeq
The coefficients $A_k$, $B_k$, $C_k$, $D_k$ are determined
by matching the poles in \eq{YonePunknown} ($P^{(1)}$ doesnot
contribute to them because $E_y(x,Y_0(x))$
has only single poles at the endpoints).
Below, we will find that $D_k=0$.

Let $k\in [1,d_2+3]$, and choose $s$ close to $e_k$:
\beq
s = e_k+\epsilon
\eeq
there must exist $\td{s}$ (unique because the potentials are
non-critical)
such that $\xx(\td{s})=\xx(s)$ and $\td{s}$ is close to $e_k$
($\td{s}$ is
the $\ss_k(x)$ in \eq{YonePunknown}):
\beq
\td{s} = e_k - \eta \virg \eta = O(\epsilon)
\eeq
By solving $\xx(s)=\xx(\td{s})$ order by order in $\epsilon$ we
get:
\beq
\eta = \l \epsilon
\virg
\l = 1 + r_k\epsilon+ r_k^2\epsilon^2 + (2r_k^3+t_k) \epsilon^3 +
O(\epsilon^4)
\eeq
where
\beq
r_k = {1\over 3} {\xx'''(e_k)\over \xx''(e_k)}
\virg
s_k = {1\over 6} {\xx^{IV}(e_k)\over \xx''(e_k)}
\virg
t_k = {1\over 60} {\xx^{V}(e_k)\over \xx''(e_k)} -r_k s_k
\eeq

From \eq{YonePunknown} we must have:
\beq
{6A_k - 2B_k \epsilon + C_k \epsilon^2 + D_k \epsilon^3}  =   -{\epsilon^4
\phi(\epsilon+\eta)\over
\xx'(e_k-\eta)  (\yy(e_k+\epsilon)-\yy(e_k-\eta))} +
O(\epsilon^4)
\eeq
We note the 3rd degree polynomial:
\beq
P_k(\epsilon) = 6A_k - 2B_k \epsilon + C_k \epsilon^2 + D_k \epsilon^3
\eeq
i.e.
\bea
P_k(\epsilon)
& = &
((1+\l)^{-2}-\zeta_1 \epsilon^2) \cr
&& (-\l \xx''(e_k) + \epsilon{\l^2\over 2} \xx'''(e_k) -
\epsilon^2{\l^3\over 6} \xx^{IV}(e_k) + \epsilon^3{\l^4\over 24}
\xx^{V}(e_k))^{-1}  \cr
&& (  (1+\l)\yy'(e_k)
+  {\epsilon\over2}(1-\l^2)\yy''(e_k)
+ {\epsilon^2\over 6}(1+\l^3)\yy'''(e_k) \cr
&& + {\epsilon^3\over 24}(1-\l^4)\yy^{IV}(e_k) )^{-1}  +
O(\epsilon^4) \cr
 & = &
- (\xx''(e_k)\yy'(e_k))^{-1}
\,\,\, (1+\l)^{-3} \l^{-1} \cr
&& (1- {3\over 2}\epsilon \l r_k
+ \epsilon^2 \l^2 s_k
- {5\over 2}\epsilon^3  (t_k+r_k s_k))^{-1}  \cr
&& (  1
+  {\epsilon\over2}(1-\l){\yy''(e_k)\over \yy'(e_k)}
+ {\epsilon^2\over 6}(1-\l+\l^2){\yy'''(e_k)\over \yy'(e_k)}
)^{-1} \cr
&& + {\zeta_1\over 2\xx''(e_k)\yy'(e_k)} \epsilon^2 + O(\epsilon^4)
\eea
It is easy to see that $t_k$ as well as $\yy^{IV}(e_k)$
disappear, and that $D_k=0$.

After a straightforward calculation, one finds:
\bea
 -8\xx''(e_k) \yy'(e_k) (6A_k - 2B_k \epsilon + C_k
\epsilon^2)
& = &
1-{1\over 3}\epsilon {\xx'''(e_k)\over \xx''(e_k)} \cr
&&  +{1\over 6}\epsilon^2\left({\xx'''(e_k)^2\over \xx''(e_k)^2}
-{\xx^{IV}(e_k)\over \xx''(e_k)} \right. \cr
& & \left. + {\xx'''(e_k)\yy''(e_k)\over \xx''(e_k)\yy'(e_k)}
 - {\yy'''(e_k)\over \yy'(e_k)}-24\zeta_1 \right)  \cr
\eea
After substitution into \eq{Yoneansatzf},
 we find the genus one correction to the resolvent:
\beq
\encadremath{
\begin{array}{lll}
Y^{(1)}(\xx(s))
& = & \sum_{k=1}^{d_2+3} {1\over 48 \xx'(s) \xx''(e_k)
\yy'(e_k)}\wei''(s-e_k) \cr
&&  + \sum_{k=1}^{d_2+3} {\xx'''(e_k) \over 48 \xx'(s)
\xx''^2(e_k)  \yy'(e_k)}\wei'(s-e_k) \cr
&& + \sum_{k=1}^{d_2+3} {
{\xx'''(e_k)^2\over \xx''(e_k)^2}
-  {\xx^{IV}(e_k)\over \xx''(e_k)}
 + {\xx'''(e_k)\over \xx''(e_k)}{\yy''(e_k)\over \yy'(e_k)}
- {\yy'''(e_k)\over \yy'(e_k)}
\over 48 \xx'(s) \xx''(e_k) \yy'(e_k)}\wei(s-e_k) \cr
&& -\sum_{k=1}^{d_2+3} {\zeta_1\over 2\xx'(s) \xx''(e_k) \yy'(e_k)}\wei(s-e_k) \cr
\end{array}
}\eeq
That function represents the partition function of a statistical
physics
model on a genus one surface with one + boundary.

Notice that
\beq
{\d F^{(1)}\over \d \epsilon} = \Gamma^{(1)} = \sum_{k=1}^{d_2+3} {
{\xx'''(e_k)^2\over \xx''(e_k)^2}
-  {\xx^{IV}(e_k)\over \xx''(e_k)}
 + {\xx'''(e_k)\over \xx''(e_k)}{\yy''(e_k)\over \yy'(e_k)}
- {\yy'''(e_k)\over \yy'(e_k)} -24 \zeta_1
\over 48 \xx''(e_k) \yy'(e_k)}
\eeq

\subsection{The free energy}

We are now going to find the free energy $F^{(1)}$ such that:
\beq
Y^{(1)}(x) = - {\d F^{(1)}\over \d V_1(x)}
\eeq

conjecture:
\beq
F^{(1)}=-{1\over 24} \ln{\left( \gamma^{4}\td\gamma^{4}
\prod_{i=1}^{d_2+3}\prod_{j=1}^{d_1+3} \left(
{\theta(e_i-\td{e}_j)\theta(2 u_\infty) \over
\theta(e_i-u_\infty)\theta(\td{e}_j+u_\infty)}
\right) \right)}
\eeq

question: how to extend that conjecture to genus $g>1$ ???

\section{Variations with respect to the potentials}

\subsubsection{Variations with $u$ fixed}

Now, we would like to compute:
\beq
\dot\xx(u) := \xx'(s){\d \xx(u)\over \d V_1(\xx(s))}
\and
\dot\yy(u) := \xx'(s){\d \yy(u)\over \d V_1(\xx(s))}
\eeq
with $u$ fixed (we will not write the $s$ dependence for lisibility in this section).

We have:
\beq
\left. {\d \yy(u) \over \d V_1(\xx(s))}\right|_{u}  =  \left.{\d
Y(\xx(u)) \over \d V_1(\xx(s))}\right|_{u}
 =  \left.{\d Y(x) \over \d V_1(\xx(s))}\right|_{x=\xx(u)}
+ Y'(\xx(u)) \left.{\d \xx(u) \over \d V_1(\xx(s))}\right|_{u}
\eeq
which implies:
\beq
\encadremath{
\dot\xx(u) \yy'(u) - \dot\yy(u) \xx'(u) =  \wei(s-u)
}
\eeq

In particular at $u=e_i$  we have:
\beq
\dot\xx(e_i) = {1\over \yy'(e_i)} \wei(s-e_i)
\eeq

We set:
\beq
\alpha_i := {\wei(s-e_i)\over \xx''(e_i)\yy'(e_i)}
\eeq

Moreover, we have:
\beq
{\dot\xx(u+1)\over \xx'(u+1)}={\dot\xx(u)\over \xx'(u)}
\virg
{\dot\xx(u+\tau)\over \xx'(u+\tau)}={\dot\xx(u)\over \xx'(u)}-\dot\tau
\eeq
where $\dot\tau := \xx'(s) {\d \tau\over \d V_1(\xx(s)) }$.
That means that the function:
\beq
f(u) := {\dot\xx(u)\over \xx'(u)} - \sum_{i=1}^{d_2+3}
\alpha_i Z(u-e_i)
\eeq
has no poles and satisfies:
\beq
f(u+1)=f(u) \virg f(u+\tau)=f(u)
- \dot\tau
+ 2i\pi \sum_i \alpha_i
\eeq
$f'(u)$ is an elliptical function with no pole, therefore it is a constant,
and $f(u)$ is a constant too.

That allows to write:
\beq
\encadremath{
\dot\xx(u) = \xx'(u) \left[ C +\sum_{i=1}^{d_2+3}
\alpha_i Z(u-e_i) \right]
}
\eeq
that implies:
\beq
\dot\tau = 2i\pi \sum_{i=1}^{d_2+3} \alpha_i
\eeq
and $C$ is a constant which is determined by the behaviours near
$\pm u_\infty$.
Indeed, consider $\dot\xx'(u)/\xx'(u)$, you get:
\beq
{2\dot{u_\infty}\over u-u_\infty} = -{2\over u-u_\infty} \left[ C
+ \sum_i \alpha_i Z(u_\infty-e_i)\right]
\eeq
and
\beq
-{(d_2+1)\dot{u_\infty}\over u-u_\infty} = -{d_2+1\over u-u_\infty} \left[ C
+ \sum_i \alpha_i Z(-u_\infty-e_i)\right]
\eeq
where $\dot{u}_\infty := \xx'(s) {\d u_\infty \over \d V_1(\xx(s)) }$.
That implies
\beq
C= -{1\over 2} \sum_i \alpha_i (Z(u_\infty-e_i) +Z(-u_\infty-e_i))
\eeq
and
\beq
-2\dot{u_\infty}= \sum_i \alpha_i (Z(u_\infty-e_i) +Z(u_\infty+e_i))
\eeq
i.e.
\beq
\encadremath{
\dot\xx(u) = {1\over 2}\xx'(u)
\sum_i \alpha_i [2Z(u-e_i)+Z(u_\infty+e_i)-Z(u_\infty-e_i)]
}\eeq

We also have:

\beq
\encadremath{
\dot\yy(u) = {1\over 2}\yy'(u)
\sum_i \alpha_i [2Z(u-e_i)+Z(u_\infty+e_i)-Z(u_\infty-e_i)]
- {\wei(s-u)\over \xx'(u)}
}\eeq

\bigskip

In particular near $u=e_i$, we find:
\beq
 \dot{e_i} = -C -\sum_{j\neq i} \alpha_j  Z(e_i-e_j)
 -{1\over 2} \alpha_i {\xx'''(e_i)\over \xx''(e_i)}
\eeq
and near $u=u_\infty$, we write:
\beq
\xx(u)\sim {A\over u-u_\infty}
\eeq
\beq
{\dot{A}\over A} = \sum_i \alpha_i \wei(u-u_\infty)
\eeq
and near $u=-u_\infty$, we write:
\beq
\yy(u)\sim {\td{A}\over u+u_\infty}
\eeq
\beq
{\dot{\td{A}}\over \td{A}} = \sum_i \alpha_i \wei(u+u_\infty)
\eeq
note that:
\beq
A=\gamma{\theta(2u_\infty)\over \theta'(0)}
\virg
\td{A}=-\td\gamma{\theta(2u_\infty)\over \theta'(0)}
\eeq

Note also that:
\beq
\xx'(s){\d \ln{\theta'(0)}\over \d V_1(\xx(s))} = {3\over 4i\pi} \dot\tau \zeta_1 = {3\over 2}\sum_i \alpha_i \zeta_1
\eeq

\subsubsection{variation of $\yy'(e_i)$}

\bea
\xx'(s){\d \left(\yy'(e_i)\right)\over \d V_1(\xx(s))}
& = & \dot\yy'(e_i) + \dot{e_i} \yy''(e_i) \cr
& = &
-\yy'(e_i)\left( \sum_{j\neq i} \alpha_j\wei(e_i-e_j) - \zeta_1\alpha_i \right)
+ {1\over 2}\alpha_i \yy'''(e_i) \cr
&& -{1\over 2} \alpha_i \yy''(e_i){\xx'''(e_i)\over \xx''(e_i)}
 + {\wei(e_i-s)\over \xx''(e_i)}\left({1\over 6}{\xx^{IV}(e_i)\over \xx''(e_i)} -{1\over 4}{\xx'''^2(e_i)\over \xx''^2(e_i)}\right) \cr
&& +{1\over 2}{\wei'(e_i-s)\xx'''(e_i)\over \xx''^2(e_i)}
 -{1\over 2}{\wei''(e_i-s)\over \xx''(e_i)}\cr
\eea
i.e.
\bea
\xx'(s){\d \left(\ln{\yy'(e_i)}\right)\over \d V_1(\xx(s))}
& = &
-\sum_{j\neq i} \alpha_j\wei(e_i-e_j) + \zeta_1\alpha_i  \cr
&& + {\alpha_i\over 2}
\left({\yy'''(e_i)\over \yy'(e_i)}+{1\over 3}{\xx^{IV}(e_i)\over \xx''(e_i)} -{1\over 2}{\xx'''^2(e_i)\over \xx''^2(e_i)}
-{\yy''(e_i)\over \yy'(e_i)}{\xx'''(e_i)\over \xx''(e_i)}\right) \cr
&& +{1\over 2}{\wei'(e_i-s)\xx'''(e_i)\over \xx''^2(e_i)\yy'(e_i)}
 -{1\over 2}{\wei''(e_i-s)\over \xx''(e_i)\yy'(e_i)}\cr
\eea

thus:
\bea
\xx'(s){\d \ln{\left(\prod_i \yy'(e_i)\right)}\over \d V_1(\xx(s))}
& = &
-\sum_i\sum_{j\neq i} \alpha_j\wei(e_i-e_j) + \zeta_1 \sum_i \alpha_i  \cr
&&
\begin{array}{r}
\displaystyle + \sum_i {\alpha_i\over 2} \left({\yy'''(e_i)\over \yy'(e_i)}+{1\over 3}{\xx^{IV}(e_i)\over \xx''(e_i)}
-{1\over 2}{\xx'''^2(e_i)\over \xx''^2(e_i)} \right.\cr
\displaystyle -\left. {\yy''(e_i)\over \yy'(e_i)}{\xx'''(e_i)\over \xx''(e_i)}\right)
\end{array}\cr
&& +{1\over 2}\sum_i{\wei'(e_i-s)\xx'''(e_i)\over \xx''^2(e_i)\yy'(e_i)}
 -{1\over 2}\sum_i{\wei''(e_i-s)\over \xx''(e_i)\yy'(e_i)}\cr
\eea

Notice that:
\beq
{\xx''(u)\over \xx'(u)} = \sum_i Z(u-e_i) - 2 Z(u-u_\infty) - (d_2+1)
Z(u+u_\infty)
\eeq
implies (expand to order $1$ in $u-e_j$):

\beq
\sum_{i\neq j} \wei(e_i-e_j) =
\zeta_1 -{1\over 3}{\xx^{IV}(e_j)\over \xx''(e_j)} + {1\over 4}{\xx'''^2(e_j)\over \xx''^2(e_j)}
+ 2\wei(e_j-u_\infty) + (d_2+1)\wei(e_j+u_\infty)
\eeq

therefore
\bea
\xx'(s){\d \ln{\left(\prod_i \yy'(e_i)\right)}\over \d V_1(\xx(s))}
& = & -\sum_j \alpha_j ( 2\wei(e_j-u_\infty) + (d_2+1)\wei(e_j+u_\infty) )  \cr
&& \begin{array}{r}
\displaystyle + \sum_i {\alpha_i\over 2}
\left( {\yy'''(e_i)\over \yy'(e_i)}+{\xx^{IV}(e_i)\over \xx''(e_i)} -{\xx'''^2(e_i)\over \xx''^2(e_i)} \right. \cr
-\displaystyle\left. {\yy''(e_i)\over \yy'(e_i)}{\xx'''(e_i)\over \xx''(e_i)}\right)
\end{array}\cr
&& +{1\over 2}\sum_i{\wei'(e_i-s)\xx'''(e_i)\over \xx''^2(e_i)\yy'(e_i)}
 -{1\over 2}\sum_i{\wei''(e_i-s)\over \xx''(e_i)\yy'(e_i)}\cr
\eea
i.e.
\bea
\xx'(s){\d \ln{\left(\prod_i \yy'(e_i)\right)}\over \d V_1(\xx(s))}
& = &
-2{\dot{A}\over A} - (d_2+1){\dot{\td{A}}\over \td{A}}  -\sum_i {\alpha_i\over 2}\zeta_1 \cr
&& \begin{array}{r}
\displaystyle + \sum_i {\alpha_i\over 2}
\left( {\yy'''(e_i)\over \yy'(e_i)}+{\xx^{IV}(e_i)\over \xx''(e_i)} -{\xx'''^2(e_i)\over \xx''^2(e_i)} \right. \cr
-\displaystyle\left. {\yy''(e_i)\over \yy'(e_i)}{\xx'''(e_i)\over \xx''(e_i)}\right)
\end{array}\cr
&& +{1\over 2}\sum_i{\wei'(e_i-s)\xx'''(e_i)\over \xx''^2(e_i)\yy'(e_i)}
 -{1\over 2}\sum_i{\wei''(e_i-s)\over \xx''(e_i)\yy'(e_i)}\cr
\eea

therefore:
\beq
F^{(1)}= K(V_2,\epsilon) -{1\over 24}
\ln{\theta'(0)^{8}\,A^{2}\, {\td{A}}^{(d_2+1)}\, \prod_{i} \yy'(e_i)}
\eeq
where $K$ may depend on $V_2$ and $\epsilon$ but not on $V_1$.
Using:
\beq
d_2 \td{g}_{d_2+1} = \gamma \td\gamma^{-d_2} \prod_i {\theta(e_i+u_\infty)\over \theta(e_i-u_\infty)}
\eeq
we can write it in a more symmetric form:
\beq
\encadremath{
F^{(1)}= K(\epsilon) -{1\over 24}
\ln{\gamma^4 \td\gamma^4 \theta'(0)^{8}
\prod_{i,j} {\theta(e_i-\td{e}_j)\theta(2u_\infty)
\over \theta(e_i-u_\infty)\theta(\td{e}_j+u_\infty)}}
}\eeq
where $K$ depends neither on $V_1$ nor on $V_2$.
$K$ may still depend on $\epsilon$.

The calculation of derivatives wrt $\epsilon$ is exactly the same, with replacing $\alpha_i$ by $1/\xx''(e_i)\yy'(e_i)$,
and we find that $K$ does not depend on $\epsilon$.

The limit where $\epsilon\to 0$ is the genus zero case, and we have $K=0$.

\section{Conclusion}
\label{conclusion}

We have computed the free energy to order $1/N^2$, in the genus one case, and it should not be too difficult
to do the calculation for all genus.
It is conjectured that $F^{(1)}$ should be the log of the determinant of the Laplacian on the algebraic curve.

\bigskip
\noindent Aknowledgements:
I am thankfull to M. Bertola, V. Kazakov, I. Kostov for fruitful discussions.

\setcounter{section}{0}
\appendix{Calculation}
\label{appendixcalcul}

We shall prove here that:
\beq\label{equation2}
\Gamma_j
= {\d F^{(0)}\over \d \epsilon_j}
\virg
\tau_{i,j} = {1\over 2i\pi} {\d^2 F^{(0)}\over \d \epsilon_i \epsilon_j}
\eeq
Notice that the second of these two equalities follows from the first.

The compatibility condition coming from \eq{WdFdV}:
\beq
{\d \over \d V_1(x)}{\d F^{(0)}\over \d \epsilon_j}
= {\d \over \d \epsilon_j}{\d F^{(0)}\over \d V_1(x)}
= {\d \over \d \epsilon_j} W_1(x)
= -{\d \over \d \epsilon_j} Y(x)
\eeq
as well as \eq{defdui} and \eq{duidGammadV}
imply that $\Gamma_j - {\d F^{(0)}\over \d \epsilon_j}$ is independent of $V_1$,
and by the same argument after an integration by parts, it must be independent of $V_2$ too.

In order to prove that $\Gamma_j - {\d F^{(0)}\over \d \epsilon_j}=0$, we can choose $V_1$ and $V_2$
such that the $x$-physical sheet contains all the $A$-cylces.
That means that the function $Y_0(x)$ has $g+1$ cuts in the physical sheet, i.e. the large $N$
average density of eigenvalues of matrix $M_1$ has a support made of $g+1$ connected intervals:
\beq
\rho(x) = {1\over 2i\pi} (Y(x+i0)-Y(x-i0))
\virg
{\rm supp}\,\rho = \bigcup_{i=1}^{g+1}\, [a_i,b_i]
\eeq
We choose the $A_i$ contours as follows:
$$
{\epsfxsize 11cm\epsffile{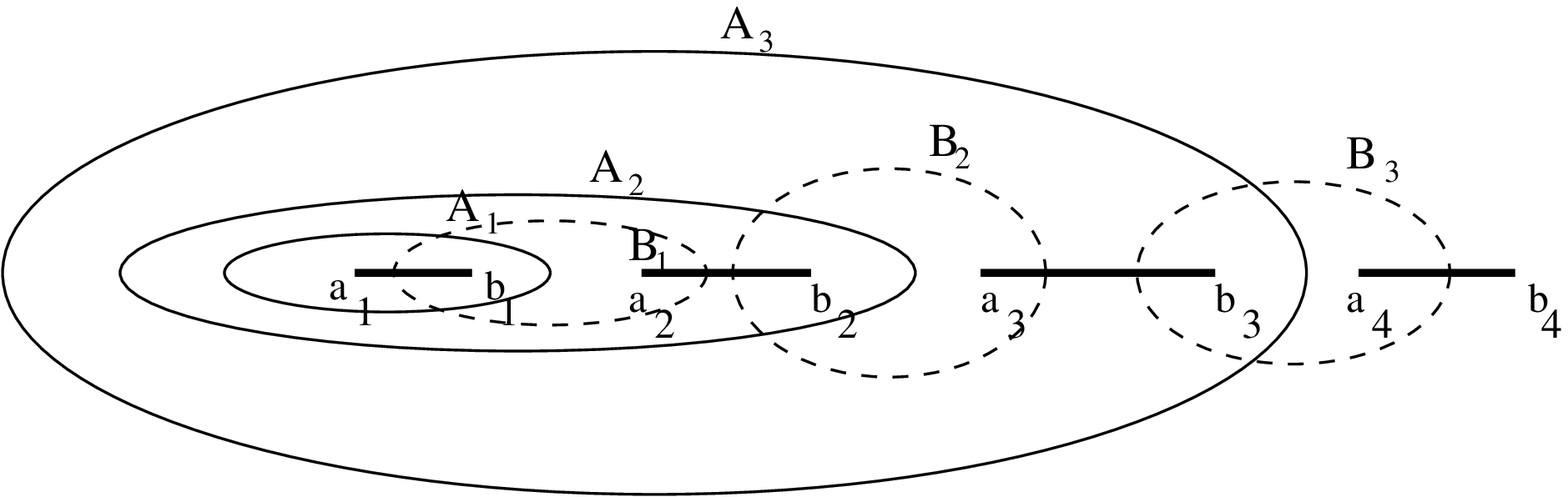}}
$$
and we can choose $V_1$ and $V_2$ such that the filling fraction of interval $[a_i,b_i]$ is:
\beq
{n_i\over N} := \int_{a_i}^{b_i} \rho(x) dx = {1\over 2i\pi} \oint_{[a_i,b_i]} y dx = \epsilon_i -\epsilon_{i-1}
\eeq
$n_i$ is the average number of eigenvalues of $M_1$ in interval $[a_i,b_i]$.

Now, following the method of \cite{David} for the one-matrix model,
compute the energy cost of moving one single eigenvalue $x$ from $a$ to
$b$, where $a$ belongs to cut $i$ and $b$ to cut $i+1$.
Basicaly: $n_i\to n_i-1$ and $n_{i+1}\to n_{i+1}+1$, i.e.
$\epsilon_i \to \epsilon_i - {1/N}$.
That energy cost can be written:
\beq
-{1\over N} {\d N^2 F\over \d \epsilon_i}
=
N \int_a^b V'_{\rm eff}(x) dx
\eeq
where the effective mean field potential $V_{\rm eff}$
experienced by one eigenvalue of $M_1$ in the presence of the potentials and the interaction with all other
eigenvalues in their equilibrium position was computed in \cite{matytsin, PZJ, BEHansatz, KazMar},
and it reduces to:
\beq
{\d F\over \d \epsilon_i}
= {1\over 2i\pi} \oint_{B_i} Y(x) dx = \Gamma_i
\eeq

\appendix{Theta functions in genus one}
\label{thetagenus1}

Consider a complex number $\tau$ such that $\Im\, \tau >0$, called the modulus.
We define the theta-function by:
\beq
\theta(u) = \sum_{n=0}^\infty \ee{i\pi \tau n^2} \sin{(2\pi n u)}
\eeq

We have:
\beq
\theta(u+1) = \theta(u)
\virg
\theta(u+\tau) = -\theta(u) \ee{-i\pi(2u+\tau)}
\virg
\theta(-u)=-\theta(u)
\eeq

We also introduce:
\beq
Z(u) = {\d \over \d u} \ln{\theta(u)}
\eeq
we have:
\beq
Z(u+1)=Z(u)
\virg
Z(u+\tau)=Z(u)-2i\pi
\virg
Z(-u)=-Z(u)
\eeq
$Z(u)$ has a single pole at $u=0$, with residue $1$:
\beq
Z(u) \sim {1\over u} + \zeta_1 u + O(u^3)
\eeq

And we introduce the Weierstrass function:
\beq
\wei(u) = -{\d\over \d u} Z(u)
\eeq
we have:
\beq
\wei(u+1)=\wei(u+\tau)=\wei(-u)=\wei(u)
\eeq
$\wei(u)$ is thus an elliptical function (doubly periodic).
It has a double pole at $u=0$:
\beq
\wei(u) \sim {1\over u^2} - \zeta_1 + O(u^2)
\eeq

Consider an elliptical function $f$ such that:
\beq
f(u+1)=f(u+\tau)=f(u)
\eeq
then:
\begin{itemize}
\item if $f$ is entire, then $f$ is a constant.
\item if $f$ is meromorphic (its singularities are poles), it must have at least
two poles (possibly a double pole).
\item if $f$ has $n$ poles $e_1,\dots,e_n$ with multiplicities $m_1,\dots,m_n$,
in the fundamental parallelogram of side $(1,\tau)$, then there exist
$m=\sum_{k=1}^n m_k$ complex numbers $A, f_1, f_2,\dots, f_{m-1}$ such that:
\beq
f(u) = A {\prod_{k=1}^m \theta(u-f_k)\over \prod_{k=1}^n \theta(u-e_k)^{m_k}}
\eeq
where we have defined $f_{m}$ such that:
\beq
\sum_{k=1}^m f_k = \sum_{k=1}^n m_k e_k
\eeq
\item An alternative representation of $f$ is the following: there exist
there exist $m+1$ numbers $A_0,A_{1,1},\dots,A_{1,m_1},\dots, A_{n,1},\dots,
A_{n,m_n}$ such that:
\beq
f(u) = A_0 + \sum_{k=1}^n \sum_{l=1}^{m_k} A_{k,l} Z^{(l-1)}(u-e_k)
\eeq
with the condition that:
\beq
\sum_{k=1}^n A_{k,1} =0
\eeq

\end{itemize}

\appendix{Theta functions arbitrary genus}
\label{thetagenusall}

\subsection{Genus $g>1$}

\beq
\theta(u) = \sum_n \ee{i\pi n^t \tau n} \ee{2i\pi n^t u}
\eeq
we have:
\beq
\theta(u+n)=\theta(u)=\theta(-u)
\eeq
and
\beq
\theta(u+\tau n) = \theta(u) \, \ee{-i\pi [2 n^t u + n^t \tau n]}
\eeq


\end{document}